\documentclass[10pt,journal,comsoc]{IEEEtran}
\usepackage{url}
\usepackage{caption}
\usepackage[noadjust]{cite}
\usepackage{graphicx}
\usepackage{multirow}
\usepackage{caption}

\usepackage{comment}
\usepackage{xspace}
\usepackage{caption}
\usepackage{subcaption}
\usepackage{makecell}
\usepackage{ragged2e}
\usepackage{rotating}
\usepackage{lscape}
\usepackage[table]{xcolor} 
\usepackage{nomencl}
\usepackage{mwe}
\makenomenclature
\usepackage{wrapfig}
\usepackage{makecell}
\usepackage{balance}
\usepackage[utf8]{inputenc}
\usepackage{ragged2e}
\usepackage{booktabs}

\usepackage{pifont}
\usepackage{tabulary}
\usepackage{comment}
\usepackage{lipsum}
\usepackage{array, tabularx, multirow, booktabs, diagbox}
\usepackage{adjustbox}

\usepackage{amsmath,amssymb,amsfonts}
\usepackage{graphicx}
\usepackage{textcomp}
\def\BibTeX{{\rm B\kern-.05em{\sc i\kern-.025em b}\kern-.08em
    T\kern-.1667em\lower.7ex\hbox{E}\kern-.125emX}}

\begin{document}
\title{Federated Learning for Digital Twin-Based Vehicular Networks: Architecture and Challenges}

\author{Latif~U.~Khan,~Ehzaz~Mustafa,~Junaid~Shuja,~Faisal~Rehman,~Kashif~Bilal,~\IEEEmembership{Senior~Member,~IEEE},~Zhu~Han,~\IEEEmembership{Fellow,~IEEE},~and~Choong~Seon~Hong,~\IEEEmembership{Senior~Member,~IEEE}
    
\IEEEcompsocitemizethanks{
\IEEEcompsocthanksitem L.~U.~Khan~and~C.~S.~Hong are with the Department of Computer Science \& Engineering, Kyung Hee University, Yongin-si 17104, South Korea.
\IEEEcompsocthanksitem E.~Mustafa, F. Rehman and K. Bilal are with the Department of Computer Science COMSATS University, Abbottabad Campus, 22044, Pakistan.
\IEEEcompsocthanksitem J. Shuja is with Department of Computer Science, National University of Computer and Emerging Sciences, Karachi, Pakistan
\IEEEcompsocthanksitem Zhu Han is with the Electrical and Computer Engineering Department, University of Houston, Houston, TX 77004 USA, and also with the Computer Science Department, University of Houston, Houston, TX 77004 USA, and the Department of Computer Science and Engineering, Kyung Hee University, South Korea.

}}

\markboth{}{}%

\maketitle






\begin{abstract} 
Emerging intelligent transportation applications, such as accident reporting, lane change assistance, collision avoidance, and infotainment, will be based on diverse requirements (e.g., latency, reliability, quality of physical experience). To fulfill such requirements, there is a significant need to deploy a digital twin-based intelligent transportation system. Although the twin-based implementation of vehicular networks can offer performance optimization. Modeling twins is a significantly challenging task. Machine learning (ML) can be a preferable solution to model such a virtual model, and specifically federated learning (FL) is a distributed learning scheme that can better preserve privacy compared to centralized ML. Although FL can offer performance enhancement, it requires careful design. Therefore, in this article, we present an overview of FL for the twin-based vehicular network. A general architecture showing FL for the twin-based vehicular network is proposed. Our proposed architecture consists of two spaces, such as twin space and a physical space. The physical space consists of all the physical entities (e.g., cars and edge servers) required for vehicular networks, whereas the twin space refers to the logical space that is used for the deployment of twins. A twin space can be implemented either using edge servers and cloud servers. We also outline a few use cases of FL for the twin-based vehicular network. Finally, the paper is concluded and an outlook on open challenges is presented.

\end{abstract}



\section{Introduction}
\setlength{\parindent}{0.7cm} Vehicular networks enable many applications (e.g., congestion control and accident reporting) for intelligent transportation systems (ITSs) \cite{noor2020survey}. Vehicular network applications are based on diverse requirements (e.g., latency and reliability) and user-defined characteristics (e.g., quality of physical experience), which are difficult to be fulfilled by the existing wireless technologies \cite{khan2022digital,khan2022metaverse}. There is a need to deploy vehicular network using two design trends, such as self-sustaining wireless systems and proactive intelligent analytics for meeting the diverse requirements. Self-sustaining wireless systems will enable to operate ITS with minimum possible intervention from the operators/users. On the other hand, proactive online learning based wireless systems will enable proactively optimizing the wireless resources for ensuring the quality of service (QoS) of various ITS applications with diverse requirements. Digital twins can be a good solution to deploy ITS applications by enabling the features of self-sustaining wireless systems and proactive online learning based systems \cite{khan2022digitalq}. \par       
Digital twins are based virtual representation of the physical systems. Additionally, optimization theory, game theory, graph theory, and machine learning (ML) can be employed by a digital twin to enable vehicular networks. For virtual modeling of the physical system, a mathematical modeling can be used which is based on several assumptions, and thus might not truly reflect the physical phenomenon. Additionally, there may be certain phenomena that cannot be modelled using mathematical modeling \cite{ali20206g}. Coping with this challenge, ML can model various challenging phenomena of vehicular networks \cite{khan2021dispersed}. ML can be based on either centralized training or distributed training. In training, centralized training transfers distributed devices data to a centralized location, and thus results in privacy leakage because vehicles are reluctant to dispense their confidential data. To tackle this limitation of centralized ML, federated learning (FL) is proposed to use distributed devices to learn a global FL model with out moving the data from devices to centralized location for training, and thus better preserves privacy compared to centralized ML. In literature, various works considered digital twins \cite{khan2022digitalq,tao2018digital,fuller2020digital}. The work in \cite{khan2022digitalq} presented the vision of digital twin-based wireless systems. Additionally, they outlined key requirements and proposed a general architecture for digital twin-based wireless systems. Tao \textit{et al.} in \cite{tao2018digital} discussed the fundamentals of digital twins. Additionally, they presented the state-of-the-art of digital twins for industries. Another work \cite{fuller2020digital} presented the concept, enablers, and open research challenges for digital twin. \par
To the best of our knowledge, different from \cite{khan2022digitalq,tao2018digital,fuller2020digital}, our work is the first one to review FL for digital twin-based vehicular networks. In contrast to the works in \cite{khan2022digitalq,tao2018digital,fuller2020digital}, our work focuses on role of FL in digital twin-based vehicular networks and present a general architecture. Also, we present an example scenario of FL towards enabling digital twin-based vehicular networks. Additionally, we present use cases of FL in twin-based vehicular networks. As such, our contributions are summarized as follows:
\begin{figure*}[!t]
	\centering
	\captionsetup{justification=centering}
	\includegraphics[width=17cm, height=10cm]{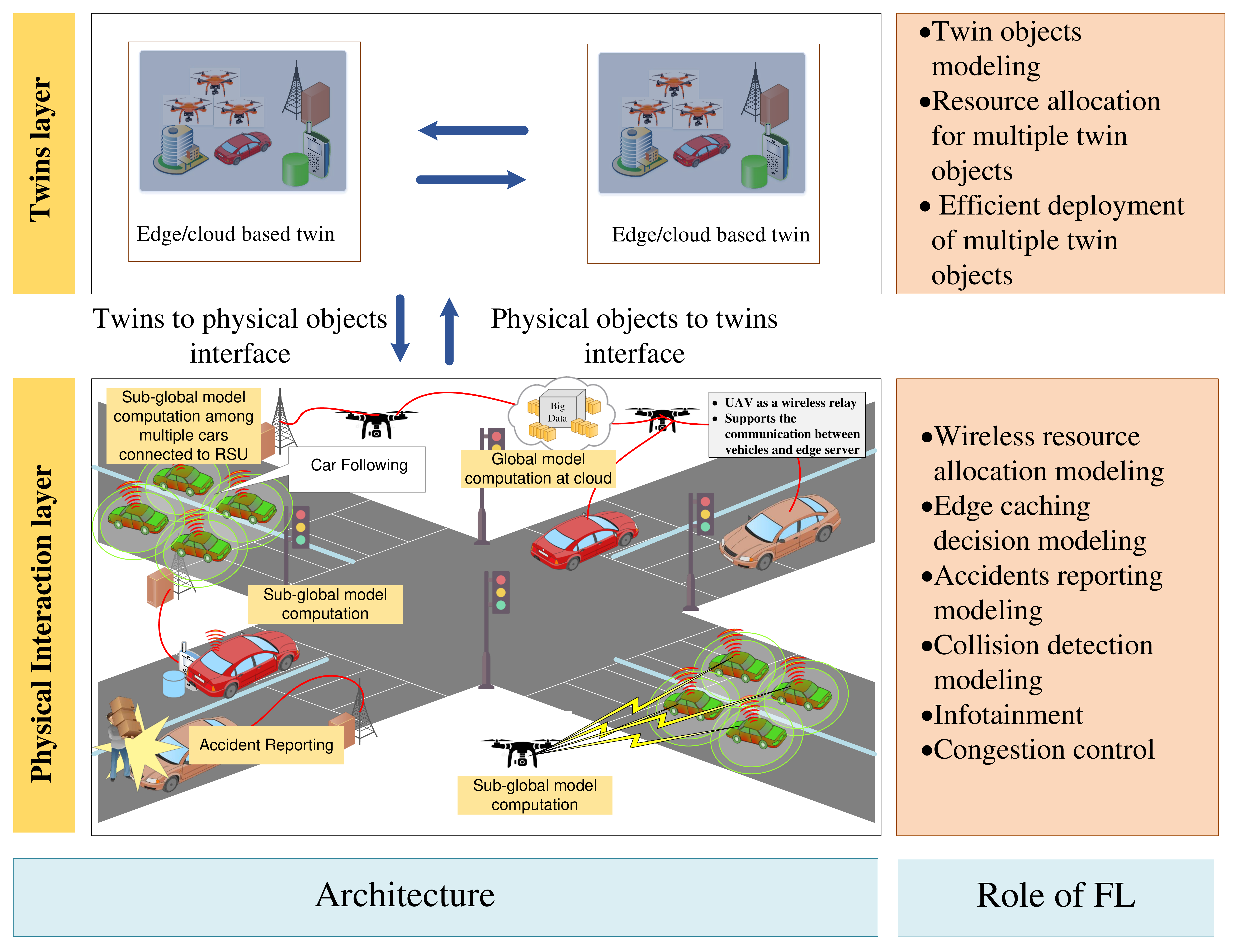}
	\caption{The architecture of federated learning for digital twin-based vehicular networks.}
	\label{fig:architecture}
\end{figure*}

\begin{itemize}
    \item We construct a general, detailed architecture of FL for digital twin-based vehicular networks. An example scenario of FL for a twin-based vehicular network is presented. Furthermore, we present the sequence diagrams for FL in twin-based vehicular networks.
    \item We present the use cases, such as intelligent analytics, edge caching, and intelligent resource management of FL towards enabling twin-based vehicular networks.
    \item Finally, we present open research challenges with their possible solutions. 
\end{itemize}

\section{Architecture of FL and Digital Twin Based Vehicular Networks}
\begin{figure*}[!t]
	\centering
	\captionsetup{justification=centering}
	\includegraphics[width=18cm, height=8cm]{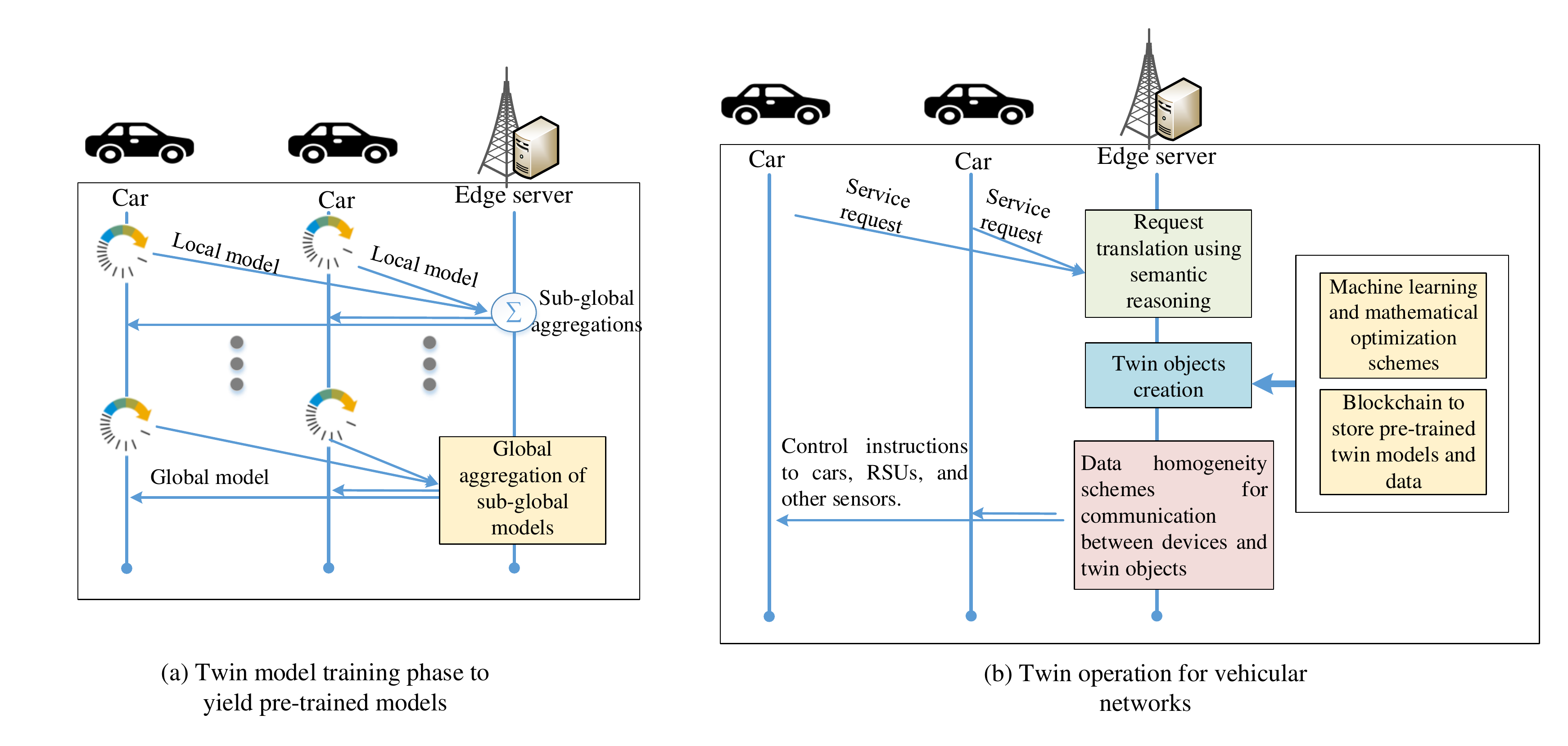}
	\caption{Sequence diagram of (a) offline twins training using DFL, and (b) online twin-based vehicular network operation}
	\label{fig:seq}
\end{figure*}

In this section, we present a high-level architecture of FL-enabled digital twin for vehicular networks, as shown in Fig.~\ref{fig:architecture}. There are two main phases in digital twin-based vehicular network: (a) offline training and (b) online operation \cite{khan2022digitalq}. In offline training, one can proactively train twin models prior to user requests, whereas online operation is based on instructing cars, sensors, and roadside units (RSUs) to serve end users, as shown in Fig.~\ref{fig:seq}. The architecture consists of two main layers: the twins layer and the physical interaction layer. The physical interaction layer consists of all entities (e.g., autonomous cars, edge-based RSUs, unmanned aerial vehicles (UAVs)) necessary for implementing digital twin-based vehicular networks. The twin layer is implemented using edge/cloud servers by using the concept of twin objects \cite{khan2022digitalq}. A twin object uses virtual representation of the physical system and such a virtual representation of the physical vehicular network can model various vehicular network functions/applications. Modeling of such network functions/applications can be performed using mathematical and experimental schemes \cite{khan2022digital}. However, mathematical modeling is mostly based on assumption, and thus might not be able to truly model the vehicular network. Meanwhile, experimental modeling might not truly model because of errors during experimentation. To address these issues, a data-driven modeling based on FL can be considered in digital twin-based vehicular networks. \par 

\begin{table*}

\caption {Comparison of various schemes for computing sub-global models for vehicular networks.} 
\label{tab:comparison} 
  \centering
  \begin{tabular}{p{4cm}p{4cm}p{4cm}p{4cm}}
    \toprule
\textbf{ }& \textbf{Management complexity}& \textbf{Communication cost}  & \textbf{Mobility management complexity} \\ \toprule

\textbf{Description}& This metric deals with the overall management of computing a sub-global model. & It deals with the communication cost for computing sub-global model. & This metric refers to address mobility of vehicles for computing a sub-global model.  \\ \midrule 


\textbf{Sub-global model for users within an autonomous car}& Low & Lowest & Lowest \\ \midrule   

\textbf{Sub-global model for multiple cars and single RSU}& High & Low & Low \\ \midrule

\textbf{Sub-global model for UAVs and cars}& High & High & High (i.e., for both mobile cars and mobile UAVs) \\

\bottomrule
\end{tabular}
\end{table*}

Although FL can offer better privacy preservation compared to centralized ML, FL has some challenges. One prominent challenge is the requirement of a large number of communication rounds for reaching a desirable convergence. Additionally, a centralized aggregation server might suffers from malfunctioning issues, which results in degradation of FL. Therefore, to improve the convergence and single point of failure in FL, a dispersed FL (DFL) can be utilized to deploy distributed aggregations in digital twin-based vehicular networks \cite{khan2021dispersed}. In DFL, first of all computation of sub-global models takes place in groups in a similar fashion as that of traditional FL. After computing sub-global model in DFL, the process of sharing sub-global models takes place, and finally global models are obtained by aggregating sub-global models, as shown in Fig.~\ref{fig:seq}b. Note that we can compute sub-global models in various ways for vehicular networks. For instance, we can compute sub-global models for enabling infotainment inside every car with multiple users. For other scenarios (e.g., traffic congestion control and lane change guidance) where data from a car is treated as a single local dataset. One can compute a sub-global model for set of cars connected to RSUs. A single RSU can be used to serve a single group for computing the sub-global model. Next, the sub-global models computed at various RSUs are aggregated to get a global model. On the other hand, if there are few RSUs and it is not possible for them to provide coverage to all autonomous cars. For a such case, one can deploy UAVs to serve cars. Multiple cars can be associated with a single UAV where sub-global model is computed. Table~\ref{tab:comparison} discusses the comparison of computing sub-global models using various ways, i.e., within a single car, multiple cars using RSU, and multiple cars using a UAV. We use management complexity, communication cost, and mobility management complexity for comparison of various schemes used to compute sub-global models for DFL in vehicular networks. The management complexity of single car-based sub-global model computation is low compared to multiple cars for RSUs and UAVs. Communication cost (i.e., access network resources) is high for UAV-based sub-global model computation and RSU-based sub-global model computation compared to single car-based sub-global model computation. The reason for highest cost for UAV-based implementation is due to the fact that UAV uses access network resources for communication between UAVs and cars. Additionally, UAVs will use access network resource for connectivity with RSUs. Mobility management is highest for UAV-based implementation due to fact that both UAVs and cars are mobile. \par
Next, we discuss where to place twin objects for vehicular networks. Twin objects are deployed at either edge or cloud depending on the computing power requirements and latency constraints. Mostly, the intelligent transportation applications (e.g., accident reporting and lane change assistance) have strict latency requirements. For such strict latency applications, it is a desirable to deploy twin objects near the devices at network edge. However, there are storage and computing power limitations at network edge. On the other hand, cloud has more storage with computing power capacity compared to edge, but at the cost of latency. Therefore, we should make a tradeoff between available computing power and latency. Another approach is to use a hybrid scheme that uses both cloud-based twin objects and edge-based twin objects. End-devices should be served (e.g., provide computing power and control signaling) by the edge-based twin objects until maximum available computing power of edge. Beyond the available computing at the edge, twin objects deployed at cloud can be used for serving the end-devices.     
\begin{figure}[!t]
	\centering
	\captionsetup{justification=centering}
	\includegraphics[width=7cm, height=5.5cm]{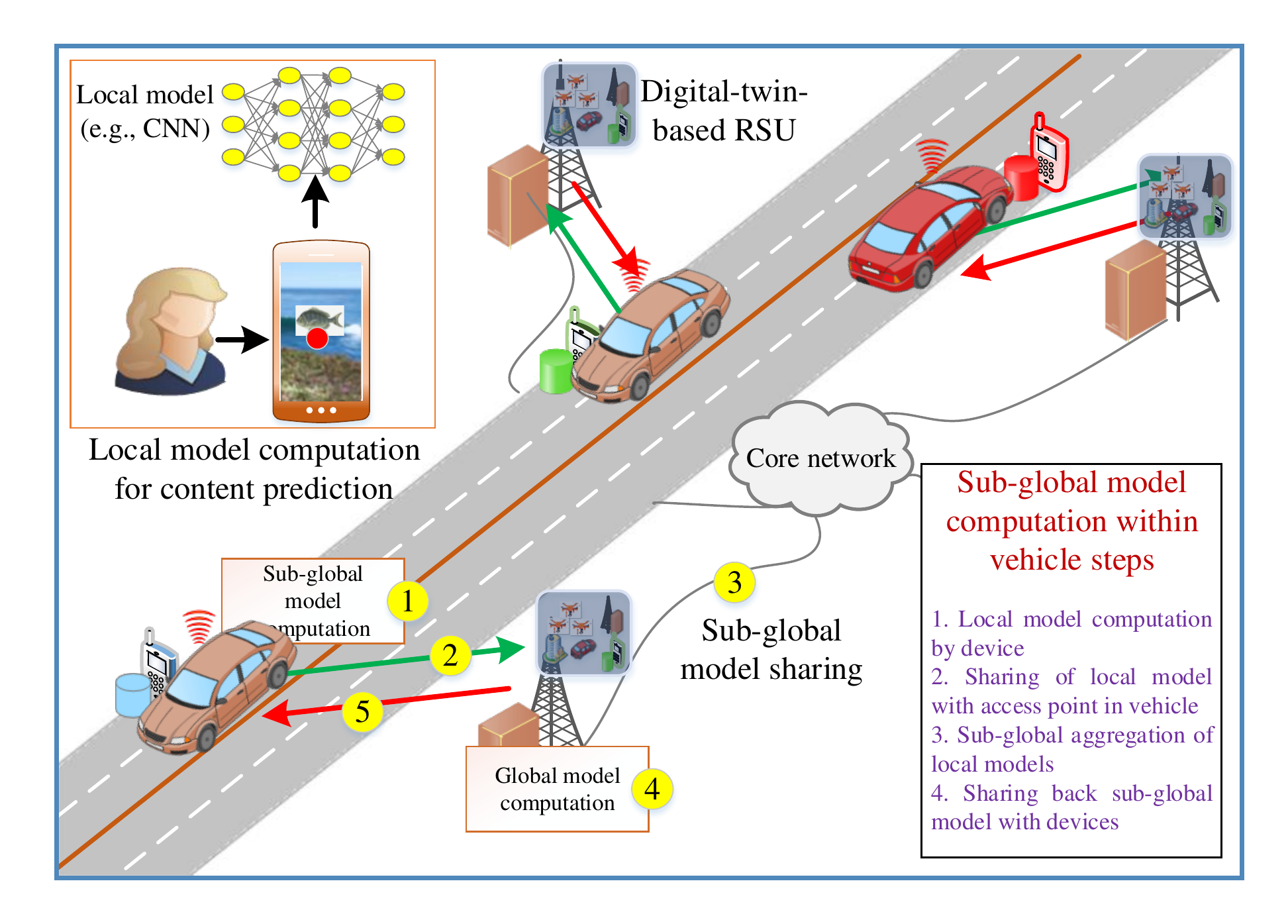}
	\caption{FL for digital twin-based infotainment in autonomous cars.}
	\label{fig:scenario}
\end{figure}

\begin{figure*}[!t]
	\centering
	\captionsetup{justification=centering}
	\includegraphics[width=17cm, height=10cm]{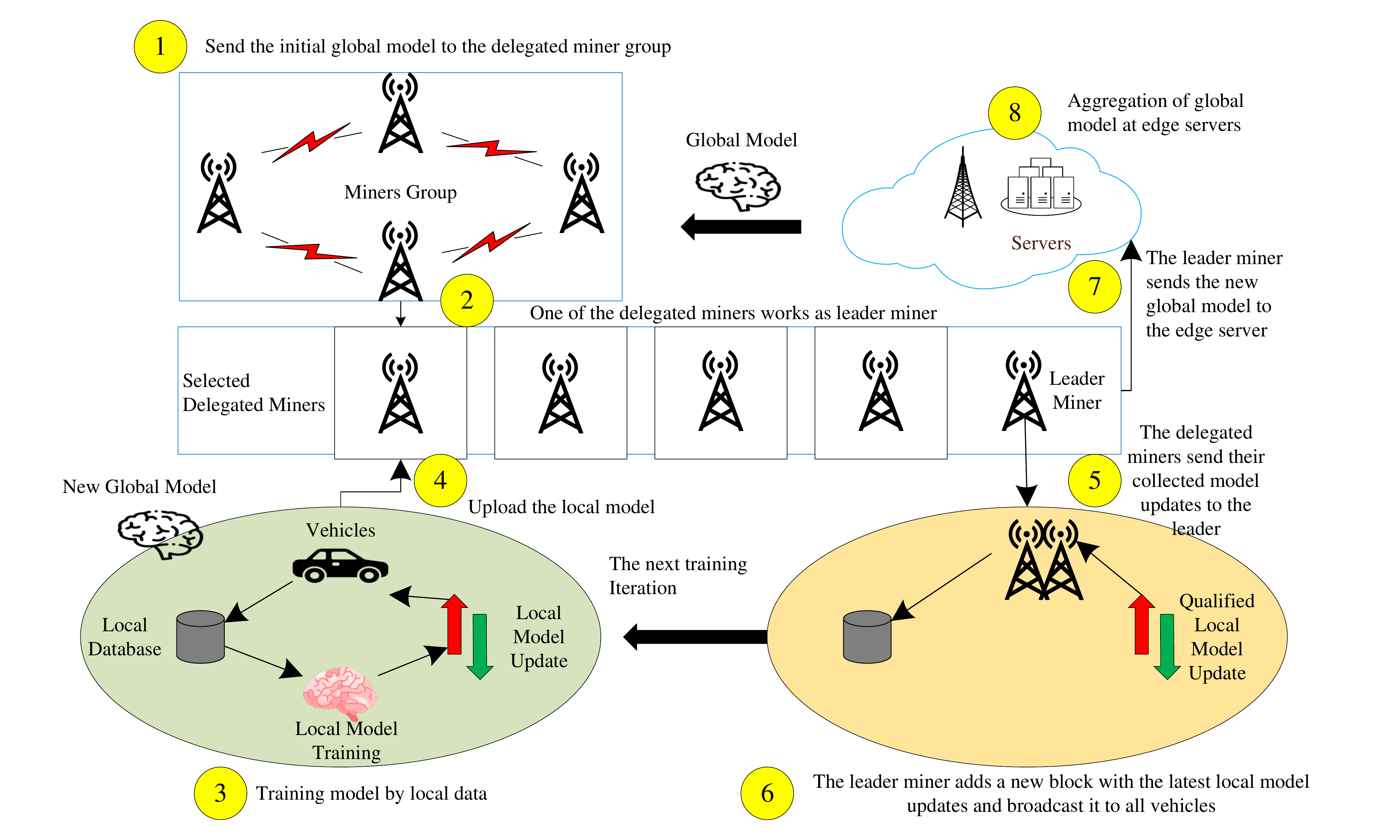}
	\caption{Blockchain-based FL for autonomous cars.}
	\label{fig:blockchain}
\end{figure*}

\subsection{Example Scenario}
We present an example scenario of digital twin-based infotainment for autonomous cars. To enable infotainment services for autonomous cars with strict latency requirements, there is a need to deploy caches at the network edge (i.e., RSUs). To do so, we deploy twin objects at the network edge for serving users. Twin objects will decide which content to cache and which to move to the cloud because edge has storage and computing power limitations compared to edge. To train such twin objects, one can use DFL, as shown in Fig.~\ref{fig:scenario}. Inside every autonomous vehicle, an access point can be installed and can act for sub-global aggregation. Inside the vehicle, cameras installed at their mobile devices can take their images that can be used by a local learning model of DFL to yield a local model. There should be certain effective local model that enable us to find the relevant infotainment item for the particular user based on passenger age, emotion, and gender. Within every autonomous vehicle, all the local models are aggregated to yield a sub-global twin model. Next to sub-global twin model computation, every autonomous vehicle can send their sub-global twin model to RSUs. Then RSUs share the sub-global twin models with each other and perform aggregations at all RSUs to yield global twin models. This process takes place iteratively until convergence. Such a twin model at the network edge will be used by twin object to decide which content to cache at particular RSUs. Additionally, twin object will control the communication and computation required for enabling infotainment in autonomous vehicles.\par

\section{Use Cases}

\subsection{Secure and Robust Federated Analytics}
Autonomous vehicles share data among each other to get more information of the environment, which improves the traffic management and reduces the risk of traffic accidents. However, sharing of data may cause serious risks to users and societies. Therefore, instead of sharing data, one can send a function of data with the edge/cloud server using federated analytics \cite{chen2021digital}. On the other hand, one can use the data generated in autonomous vehicles to train various ML models. Training based on centralized ML for traffic prediction and network management contributes to the privacy issues. Such a privacy issue is due to transferring the devices data to the remote cloud for training. FL effectively reduces these privacy concerns normally occurring in centralized ML based training, and during sharing of data. During the FL process, vehicles forward their local models to a centralized model aggregator. Such an aggregator might suffer from security attacks and malfunction due to a physical damage. To address this issue, one can use a consortium blockchain based FL framework to enable decentralized, secure, and reliable FL without a need of centralized model aggregator. See Fig.~\ref{fig:blockchain} for detailed consortium blockchain based FL for ITS. The model updates from dispersed vehicles are substantiated by miners to block unreliable model updates, and then are stored on the blockchain. Although FL can better preserve privacy, it has some privacy concerns. An important information can be inferred from the local model updates by a malicious aggregation server. Therefore, to resolve this issue, we can use a differential privacy that adds a noise to local model before transmitting to the global server \cite{qi2021privacy}.  \par

\begin{figure*}[!t]
	\centering
	\captionsetup{justification=centering}
	\includegraphics[width=16cm, height=6cm]{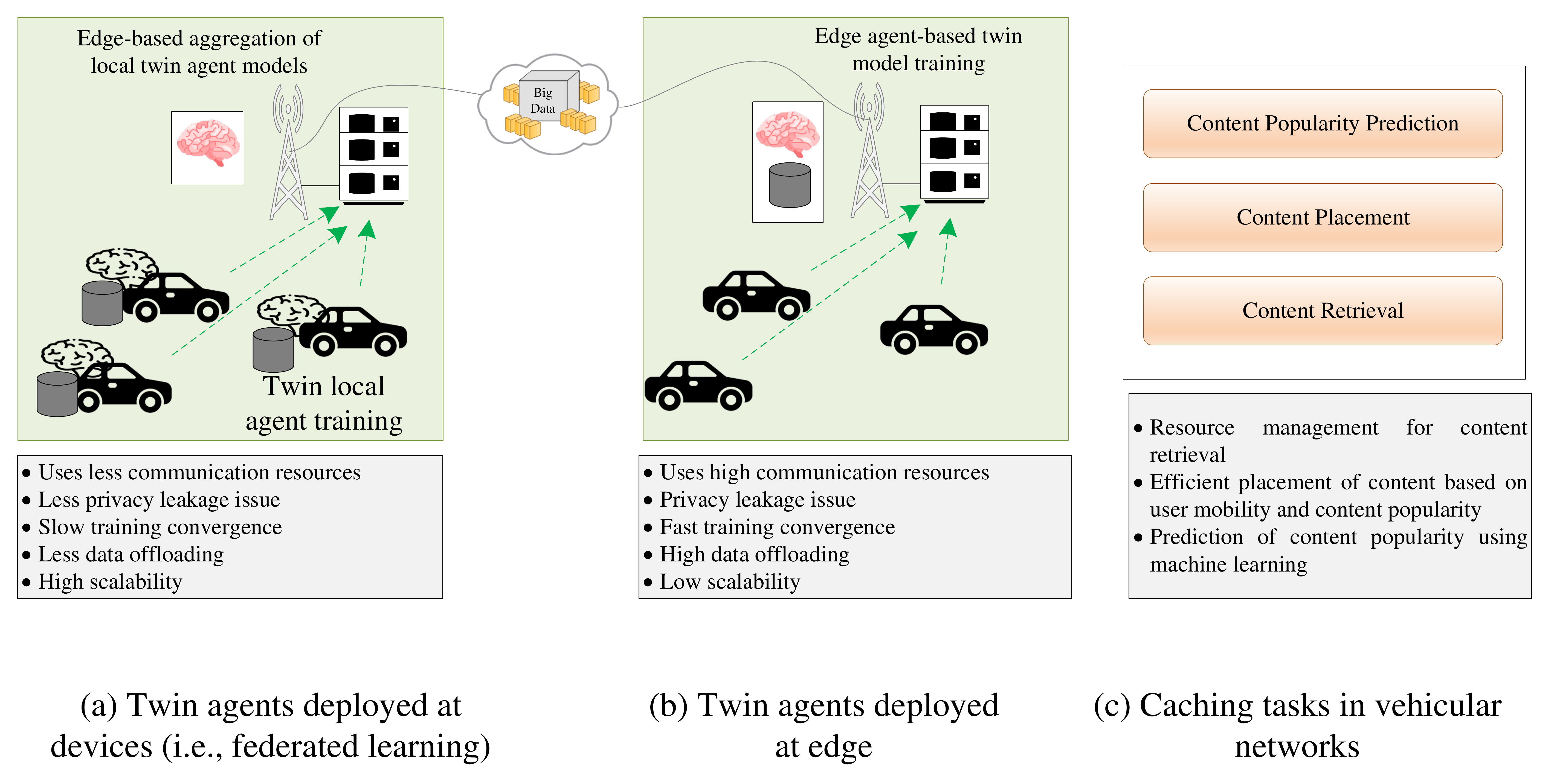}
	\caption{Deep reinforcement learning-based on FL for caching in vehicular networks.}
	\label{fig:cache}
\end{figure*}

\subsection{Intelligent Edge Caching}
A massive number of autonomous cars will request for contents from a remote cloud that will cause an increase in service requests on a vehicular network. Fetching contents from a remote cloud in ITS for various applications will suffer from high latency. To address the escalating content requests, edge-based caching can be used that reduces service latency and network traffic in vehicular networks. However, mobility in vehicular networks setup highly influences the prediction and content popularity. Vehicles connected to the edge server keeps moving in transportation systems making the cached contents to be out of date. The caching process has three major tasks: content placement in the cache, content popularity prediction, and content retrieval \cite{javed2021ai}. To deploy caching at the network edge, one can use twins deployed at edge that will perform all the three tasks of caching. In twins, the contents are stored at the edge based on their popularity. Upon request from the users, twin will provide user with the content (i.e., content retrieval) in an efficient way by performing effective resource allocation. For resource allocation, the twin uses ML, optimization theory, game theory, and graph theory. There can be many ways for predicting the content prior to storage at edge by twins. ML based content popularity prediction effectively works in mobility scenario. However, these schemes-based on centralized training need to upload and centrally process user data at a central server. Autonomous cars frequently generate data whose transfer to the centralized server is challenging. To address this limitation, one can use deep reinforcement learning (DRL)-based on FL. Fig.~\ref{fig:cache} illustrates the DRL for twins-enabled intelligent caching. One can deploy twins (i.e., DRL agent) either at the network edge or devices. Twin agent deployed at devices can perform training followed by transmitting the trained twin model to the edge for aggregation to yield a global twin model. Although this approach can send only updates to the edge server, and thus can better preserve privacy compared to centralized training of twin agents, mostly the end-devices have computing power (i.e., CPU-cycles/sec) constraints. To address this challenge, devices can send their data to the edge for training a twin agent. Although this approach can easily train twin agent; however, at the cost of a slight loss in privacy due to moving of devices data to the edge server. Therefore, a tradeoff needs to be made between performance and privacy preservation. 

\subsection{Intelligent Resource Management}
There are two main aspects of resource management for computing and communication resources in digital twin-based vehicular networks: (a) resource management for twin signaling and (b) resource management for ITS. Twin signaling is necessary to carry out the control signals for a twin-based system, whereas resource management for applications. The allocation and management of computational and communication resources are complex and challenging goals in vehicular edge computing. To efficiently manage computing and wireless resources, one can efficiently use twins. A twin may use optimization theory for performing resource management. However, there are some resource management functions in intelligent transportation system, that can not be well modeled using mathematical optimization. FL can be preferably used to tackle this issue. Twin based on FL will enable vehicular networks with efficient management of network resources by offline analysis and online control. In offline analysis, autonomous vehicles/devices trains local models and shared with the twin deployed at the edge for global aggregation to yield a global twin model. This process will occur iteratively to train a global twin model. A blockchain network can used to store these pre-trained model for future use. The twin will retrieve the pre-trained model to serve the users via efficient resource management. The main reason for performance improvement using digital twin for vehicular networks will be due to its feature of enabling proactive intelligent analytics \cite{khan2022digitalq}. Such analytics will offer us with the proactive analysis prior to user requests. Additionally, twin-based implementation will migrate the vehicular networks towards the self-sustaining wireless systems that are one of the key design aspect of the $6$G wireless systems \cite{saad2019vision}. Such self-sustaining vehicular networks will enable various applications/functions with minimum possible intervention from the network operators/end-users.\par

\section{Open Challenges}
\subsection{Mobility-Aware Association of Autonomous Vehicles}
{\em How does one manage the mobility-aware association of autonomous cars for FL to learn twin models in vehicular networks?} In a digital twin-based vehicular network, autonomous vehicles have high mobility, and thus may be difficult to get seamless connectivity with RSUs. On the other hand, seamless connectivity between vehicles and RSUs is required during FL. Therefore, we must effectively manage the mobility of vehicles. To address this challenge, one can use mobility management schemes based on deep learning. Such a deep learning-based mobility management scheme will predict the future locations of mobile vehicles, and thus helps associate the sets of vehicles to RSUs for FL (i.e., more specifically groups in DFL) that can remain seamlessly connected for more time compared to other vehicles.\par   

\subsection{Vehicular Twin Objects Migration}
{\em How do we perform efficient migration of FL-based twin objects in high mobility vehicular networks?} Twin objects serving a set of vehicles might not be able to get seamless connectivity due to the high mobility of autonomous cars. Therefore, one must migrate the twin objects deployed in the twin layer. Similar to a mobility-aware association, we can propose mobility management schemes to enable efficient and effective migration of the twin objects as per the mobility of the devices. Such a mobility management scheme can use deep learning to predict future locations of vehicles, and thus can efficiently migrate twin objects to relevant edge/cloud servers.

\subsection{Fairness-Enabled FL for Twin-Based Vehicular Networks}
{\em How do we enable fairness-enabled FL for training twin models in high mobility vehicular networks?} In vehicular networks, due to increased mobility and wireless channel degradation, needs to be made cars will experience poor performance compared to other cars. Additionally, due to locally available datasets at cars, some cars might not perform well. Such poor-performing cars will less influence the global FL model compared to better (i.e., both local model and wireless channel) performing nodes. Therefore, there will be fairness issues that need to be resolved. Fairness issues due to a wireless channel can be resolved by applying fairness-enabled association and resource allocation schemes. On the other hand, fairness issues due to learning algorithms can be resolved by applying fairness-based FL schemes \cite{ezzeldin2021fairfed}. \par             
\subsection{Personalized Twin Models}
{\em How does one train a twin model based on FL that can effectively model particular end-user functions/behavior?} Generally, training a twin model based on FL results in the general model that may not fit the particular end-device twin model. However, getting a global twin model using FL can be used for getting a personalized device FL-based twin model. Such a personalized twin model can be easily obtained with less efforts for further training at end-devices. To do so, one can use a Model-Agnostic Meta-Learning (MAML) \cite{fallah2020personalized}. The MAML-based implementation finds an initial point (i.e. initial point derived in distributed manner) that is shared among all end-devices. Every device then performs well after updating the given model by using few steps of a gradient-based method.

\par
\subsection{Scalable Twin Models}
{\em How do we enable massive number of vehicles/vehicular devices to effectively train twin models deployed at network edge/cloud?} Enabling a massive number of cars/devices in cars to participate in training of a twin model requires a significant amount of communication resources and computation resources as well as seamless communication. Computing and communication resources are need for local training and transferring learning model updates between devices and aggregation server, respectively. For a seamless communication, one should properly manage mobility of cars. On the other hand, for efficient communication one must propose schemes that can effectively manage communication resources. To do so, one can use heuristic schemes. Although heuristic schemes check all possible options, and thus provides better results but at the cost of high computational complexity. To resolve this issue, one can use decomposition-relaxation-based schemes. However, these schemes will suffer from approximation errors. To address the limitations of the aforementioned challenges, one can use matching theory-based schemes.  

\section{Concussions}
In this article, we have presented the role of FL towards enabling digital twin-based vehicular networks. A general architecture with role of FL in enabling digital twin-based vehicular networks is also proposed. Furthermore, use cases of FL for digital twin-based vehicular network are also outlined. Finally, open challenges with their causes and possible solutions are presented. We concluded that FL is a promising candidate to enable a digital twin-based vehicular network.

\bibliographystyle{IEEEtran}
\bibliography{database2}

\begin{IEEEbiography}[{\includegraphics[width=1in,height=1.25in,clip,keepaspectratio]{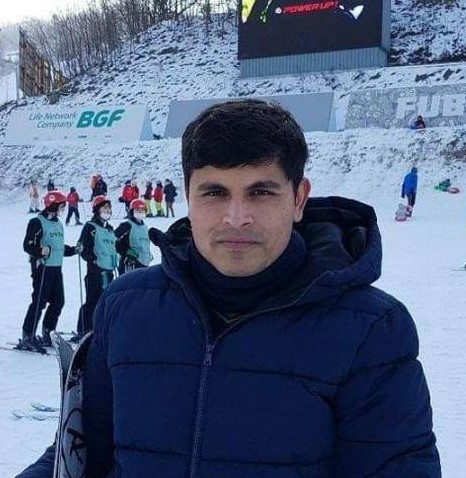}}]{Latif U. Khan} received his Ph.D. degree in Computer Engineering at Kyung Hee University (KHU), South Korea. Prior to that, He received his MS (Electrical Engineering) degree with distinction from University of Engineering and Technology, Peshawar, Pakistan in 2017. His research interests include analytical techniques of optimization and game theory to edge computing and end-to-end network slicing. 
\end{IEEEbiography}

\begin{IEEEbiography}[{\includegraphics[width=1in,height=1.25in,clip,keepaspectratio]{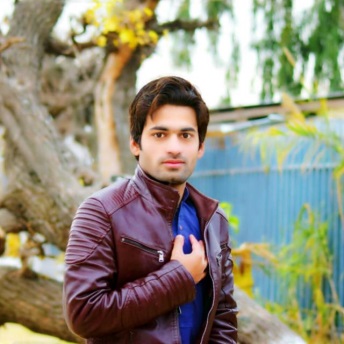}}]{Ehzaz Mustafa} is a Ph.D. student at COMSATS University Islamabad, Abbottabad Campus, Pakistan. He is also serving as lecturer at Government College Abbottabad. His research interest include Mobile Edge Computing, Machine learning, and Wireless networks. 
\end{IEEEbiography}

\begin{IEEEbiography}[{\includegraphics[width=1in,height=1.25in,clip,keepaspectratio]{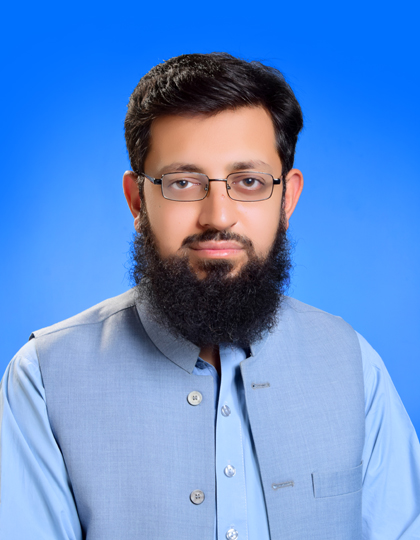}}]{Junaid Shuja} completed his Ph.D. from University of Malaya, Malaysia in 2017 under the BrightSpark scholarship program. He received “Graduate on Time” award from University of Malaya. He is an Assistant Professor at CUI, Abbottabad Campus, Pakistan. His research interests include application of machine learning techniques in edge computing, energy efficient cloud data centers, and mobile cloud computing. He has published research in more than 50 International journals and conferences.
\end{IEEEbiography}

\begin{IEEEbiography}[{\includegraphics[width=1in,height=1.25in,clip,keepaspectratio]{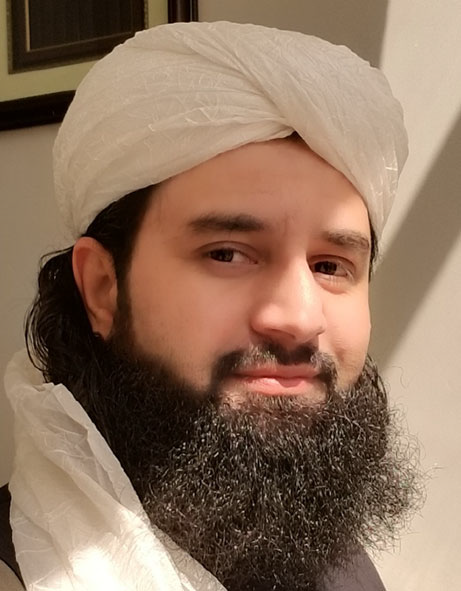}}]{Kashif Bilal} received his PhD from North Dakota State University USA. He is currently working as Associate Professor at COMSATS University Islamabad, Pakistan. His research interests include cloud
computing, energy efficient high speed networks, UAVs communications, and crowdsourced multimedia. Kashif was awarded CoE Student Researcher of the year 2014 based on his research contributions during his doctoral studies at North Dakota State University.
\end{IEEEbiography}

\begin{IEEEbiography}[{\includegraphics[width=1in,height=1.25in,clip,keepaspectratio]{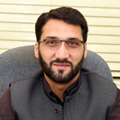}}]{Faisal Rehman} received the M.S. and Ph.D.
degrees in computer science from COMSATS University Islamabad at Abbottabad, Pakistan, in 2010 and 2018, respectively. He is currently an Assistant
Professor with COMSATS University Islamabad. His research interests include recommender systems, green computing, and wired and wireless
networks.
\end{IEEEbiography}

\begin{IEEEbiography}[{\includegraphics[width=1in,height=1.25in,clip,keepaspectratio]{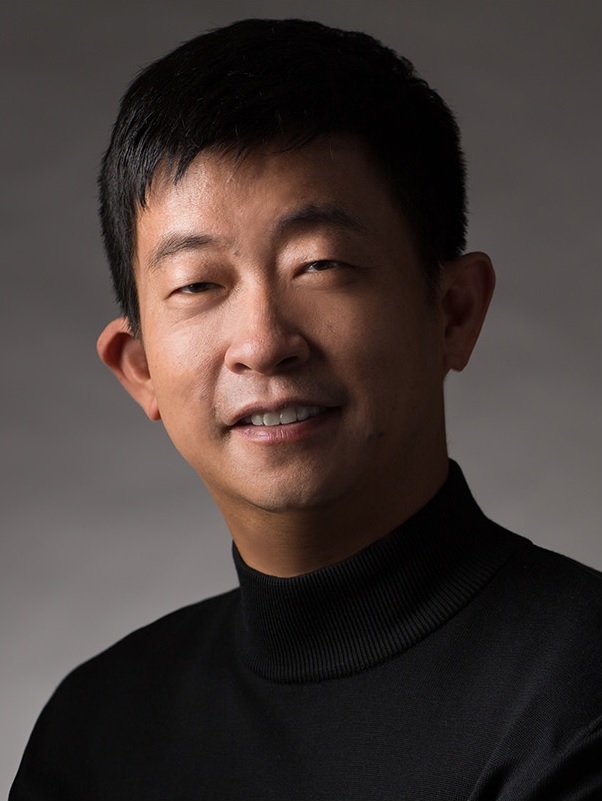}}]{Zhu Han}(S’01, M’04, SM’09, F’14) received his Ph.D. degree in electrical and computer engineering from the University of Maryland, College Park. Currently, he is a professor in the Electrical and Computer Engineering Department as well as in the Computer Science Department at the University of Houston, Texas. Dr. Han is an AAAS fellow since 2019. Dr. Han is 1$\%$ highly cited researcher since 2017 according to Web of Science.
\end{IEEEbiography}

\begin{IEEEbiography}[{\includegraphics[width=1in,height=1.25in,clip,keepaspectratio]{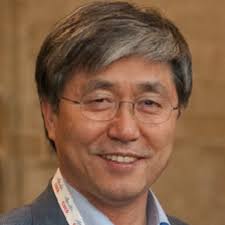}}]{Choong Seon Hong} (S’95-M’97-SM’11) is working as a professor with the Department of Computer Science and Engineering, Kyung Hee University. His research interests include future Internet, ad hoc networks, network management, and network security. He was an Associate Editor of the IEEE Transactions on Network and Service Management, Journal of Communications and Networks and an Associate Technical Editor of the IEEE Communications Magazine.
\end{IEEEbiography}

\end{document}